\documentclass[a4paper]{article}

\usepackage[english]{babel}
\usepackage[utf8x]{inputenc}
\usepackage{amsmath}

\usepackage{graphicx}
\usepackage{physics}
\usepackage[colorinlistoftodos]{todonotes}

\title{Building a bridge between Classical and Quantum Mechanics}
\author{Peter Renkel}

\begin{document}
\maketitle

\begin{abstract}
The way Quantum Mechanics (QM) is introduced to people used to Classical Mechanics (CM) is by a complete change of the general methodology (e.g. \cite{bib:Landau}) despite QM historically stemming from CM as a means to explain experimental results. Therefore, it is desirable to build a bridge from CM to QM. 

This paper presents a generalization of CM to QM. It starts from the generalization of a point-like object and naturally arrives at the quantum state vector of quantum systems in the complex valued Hilbert space, its time evolution and quantum representation of a measurement apparatus of any size. Each time, when generalization is performed, there is a possibility to develop new theory giving up most simple generalizations. It is shown that a measurement apparatus is a special case of a general quantum object. An example of a measurement apparatus of an intermediate size is considered in the end.
\end{abstract}

\section{Introduction}

The basic problem with how QM is introduced is the fact that new definitions and relations between them are presented without keeping strong connection with CM.
Indeed, QM is a more general theory than CM, and the latter can be obtained in the limit of the former. However, we live and interact with the classical world and our measurement apparatuses are mostly classical. Therefore, it is important to understand how we build QM starting from CM and generalizing it. Generalization is often not unique so the generalization we choose should be shown and justified.

Another problem related to the above is how measurement is introduced. In spite of the fact that it is central to QM, usually one loosely defines an apparatus as a macroscopic quantum object, and without clarifying what one means by 'macro' and why it is important, one jumps into Quantum Postulate \cite{bib:Copenhagen}. This postulate contradicts the unitarity of the Evolution Operator, and hence all kind of paradoxes like 'Schr\"{o}dinger's cat paradox' arise and give rise to multiple interpretations.

In this paper, I try to build the bridge between CM and QM. I obtain each QM postulate or a definition as a generalization of existing ideas in CM. Then I obtain QM parameters by matching CM to QM.  For objects in QM that don't have corresponding matches in CM, I show the way of how to arrive at them. I use Occam's razor \cite{bib:Occam} to choose the simplest generalization if several generalizations are possible. If I don't know how to avoid arbitrariness in a specific case, I try to explicitly pose a question. I give a special attention to the Problem of Measurement as central to QM. I try to avoid arbitrariness of the Quantum Postulate by explaining what is an apparatus, how to arrive at its definition and why it plays a special role in a bridge between QM and CM. I also try to avoid interpretations and present only the facts that follow from generalization of the CM.

\section{Basics}
\label{sec:basics}
The discussion below equally well applies to point-like objects and systems, characterized by generalized coordinates and momenta, or generalized coordinates and first derivatives of the coordinates. But starting from coordinate representation will give us an opportunity to easier map the actual dynamics to reality. 
\begin{enumerate}
\item In CM we start with a notion of a point-like object. In CM, it is characterized by two vectors $\{\vec{x}, \vec{p}\}$. There is seemingly some arbitrariness in the number of vectors characterizing the point-like object. Why do we need to use two vectors and not three or one? 
\item  In QM we generalize this notion by saying that a point-like object is smeared over the whole space. Since we greatly increased the number of degrees of freedom for the object, let us see if we still need momentum as an additional degree of freedom\footnote{Therefore, we won't utilize the Wigner distribution \cite{bib:Wigner}}. Thus, we start by providing the "density" of such an object in point $\vec{x}$ as $\rho(\vec{x})$\footnote{It has nothing to do with the density matrix used in quantum statistics.}, where $\rho(\vec{x})>0$. The integral of such a function over the whole space is 1 corresponding to 1 object: 
\begin{equation}
\int \rho(\vec{x}) d^3x= 1 \label{eq:density}
\end{equation}
In CM, such a density is always a delta function of coordinates centered around the point where this object is located\footnote{As we will show, the density does not provide all information, but further generalizations will yield a consistent theory}.
\item In CM, since $\vec{x}$ is a vector in a $R^3$ space, its coordinates will change depending on the system of coordinates, the orts are given by the axes of the system of coordinates.
\item In QM, our density $\rho(\vec{x})$ is a real-valued function. Therefore, we can try to represent this function as a vector in the real Hilbert space, where the basis is given by a particular set of orthonormal functions (e.g. delta functions). The coordinates (or coefficients of expansion) will change depending on the choice of the basis functions. Considering vectors, we have to define an invariant - a number that doesn't change when changing system of coordinates. The obvious candidate is vector's norm or the sum of squares of the coefficients. What is the physical quantity for an object that should be invariant? The simplest candidate is the number of objects. Therefore, we state that our candidate vector for one object is normalized and has norm 1 corresponding to one object. But stating that $\rho(\vec{x}) = \sum C_i \rho_i(\vec{x})$, where $\rho_i(\vec{x})$ are basis functions, $C_i$ are real coefficients of expansion, we need to also require that $
\int \rho^2(\vec{x}) d^3 x= 1$ which contradicts equation \ref{eq:density}. Looking again at the equation \ref{eq:density}, it is easy to see that introducing a new function $d(\vec{x}): R^3 \to R$ such that:
\begin{equation}
\rho(\vec{x}) = d^2(\vec{x}), \label{eq:distribution}
\end{equation}
with:
\begin{equation}
d(\vec{x}) = \sum C_i d_i(\vec{x}),
\end{equation}
where $d_i(\vec{x})$ are orthonormal basis functions, $C_i$ are the coefficients of expansion, we satisfy our requirements. We will call this function - distribution function.

In conclusion, to be able to transform basis functions while keeping the number of objects invariant, we should move from density $\rho(\vec{x})$ to a new function - distribution function defined by equation \ref{eq:distribution}. Since distribution is a vector in the Hilbert space, we can completely get rid of index $\vec{x}$. Therefore, we are now associating a distribution state vector $\ket{D}$ in real Hilbert space with a particle. 

For an arbitrary expansion:
\begin{equation}
\ket{D} = C_1 \ket{D_1} + C_2 \ket{D_2},
\end{equation}
where $\ket{D_1}$ and $\ket{D_2}$ are two normal vectors, not necessarily orthogonal to each other:
\begin{equation}
1 = \bra{D}\ket{D} = C_1^2 + C_2^2 + 2 C_1 C_2 \bra{D_1}\ket{D_2},
\end{equation}
where $\bra{D_1}\ket{D_2} = \int d_1(\vec{x}) d_2(\vec{x}) d^3 x$ is a scalar product of the two vectors.
\end{enumerate}

\section{Time Evolution}
\label{sec:evolution}
\begin{itemize}
\item In CM, the coordinates of a particle can be described by $\vec{x}(t)$ - a \textbf{function} of $t$. Also, there is a function that takes coordinates at time $t_0$ and gives us new coordinates at time $t$.
\item In QM, the $d(\vec{x}_0, t)$ is a function of $t$ for a given $\vec{x}_0$. Consequently, there is an operator acting on initial function (or a vector in Hilbert space) at $t_0$ to give us a function (or a vector in Hilbert space) at time $t$. It is not easy to generalize CM Newtonian laws for such a case. So let us start by stating the minimal set of meaningful requirements for such an operator. We will use Occam's razor, so if we could impose a range of constraints, we use the simplest. After stating such requirements and finding an operator satisfying them, we should make sure that we can get CM in the limit when above mentioned densities are delta functions.
\begin{enumerate}
\item This operator preserves the norm (we stated that our $d(\vec{x})$ is always normalized).
\item This operator should be linear. This is the simplest non-trivial property of operators we may impose.
\end{enumerate}
\item The simplest non-trivial linear transformation that preserves the norm is just a rotation - an orthogonal transformation. To derive the exact form of this operator, we first need to investigate how two quantum systems interact with each other. We will do it in Section \ref{sec:measurement}.
\end{itemize}

\section{A particle (system) with many independent degrees of freedom}
\begin{enumerate}
\item In CM, a particle or a system with many independent degrees of freedom is characterized by a set of numbers. Each number corresponds to one degree of freedom.
\item In QM, by analogy we claim that a particle or a system is smeared over a $R^M \times H$ dimensional space, where $M$ is the number of degrees of freedom. Thus, we define a vector for a particle in a specific point of this $R^M\times H$ dimensional space and get a $\rho(a_1, a_2, ..., a_M)$, where $a_i$ are independent characteristic parameters of a particle.
\end{enumerate}

\section{Multiple particles}
\label{sec:multiple_particles}
\begin{enumerate}
\item Suppose we have $N$ \textbf{non-interacting} particles. In CM, the generalization is straightforward: we have $N$ vectors in $R^6$ or 1 vector in $R^{6\times N}$ space\footnote{Because in CM, we need to provide both coordinates and velocities, while in QM, it is enough to provide the state vector $\vec{x}_1...\vec{x}_N$}. 
\begin{itemize}
\item Generalization for QM is also straightforward. If we are given $N$ non-interacting particles, we can consider each of them independently of all others. The state vector of the system is just a collection of respective state vectors\footnote{Generally speaking, the set of all such collections doesn't constitute a linear space}. To describe the system, where $i^{th}$ particle has $d_i$ degrees of freedom, we need $\sum_N d_i$ degrees of freedom. For example, for a two-particle state:
\begin{equation}
\rho(\vec{x}_1, \vec{x}_2) = \rho_1(\vec{x}_1) \rho_2(\vec{x}_2), \label{eq:two_particles}
\end{equation}
where $\rho_1(\vec{x}_1)$ and $\rho_2(\vec{x}_2)$ are the densities of the first and second particles respectively. There are two different density functions here and two different coordinates. The reason equation \ref{eq:two_particles} takes this form is because the particles don't interact, as we argued above, our evolution operator is linear, and therefore we want any evolution of the first particle to not effect the second particle\footnote{The only thing it does is it multiplies the vector corresponding to the second particle with a constant}. In other words, any operator representing evolution of the first particles will act only on the density corresponding to the first particle\footnote{The density of the second particle will multiply by a number, but since densities are normalized, it doesn't change anything.}. Corresponding distributions are:
\begin{equation}
d(\vec{x}_1, \vec{x}_2) = \pm d(\vec{x}_1) d(\vec{x}_2). \label{eq:two_particles_distributions}
\end{equation}
\end{itemize}
The sign in equation \ref{eq:two_particles_distributions} doesn't change with time and unphysical as long as the particles don't interact.
\item Suppose we have $N$ \textbf{interacting} particles. In CM, the system is described by $N$ vectors, time evolution of each of them depends on the positions and momenta of the others. We can also describe the system with 1 vector in $R^{6 \times N}$ space corresponding to $N$ particles but the evolution of this vector is complicated. 
\begin{itemize}
\item If we try to naively generalize this understanding to QM, we run into a problem. Since the particles are now interacting, the fact that we know the density distribution for one of them gives us some information about the others. Say for an attractive force, we expect the densities of the particles be peeked roughly in the same area. Hence there is a correlation between the densities of particles. 

Let's consider a system of two interacting particles. Suppose, they both have distributions:
\begin{equation}
d(x) = C_1 \exp\left(-\frac{(x - x^1_0)^2}{2\sigma^2}\right) + C_2 \exp\left(-\frac{(x - x^2_0)^2}{2\sigma^2}\right),
\end{equation}
where $C_1$ and $C_2$ are some normalization coefficients, $x^1_0$ and $x^2_0$ correspond to peeks and $\sigma$ correspond to widths of the two Gaussians\footnote{For simplicity, we assume the Gaussians are symmetric in x-y plane.}.
But we expect the distribution of the second particle to peek at the same places where the distribution of the first particle peeks (points $x^1_0$ and $x^2_0$), otherwise we will lose the information about correlations. Therefore, it is not enough to know the distributions of each particle, we also need to provide their correlations.

The solution is to introduce the distribution of the pair:
\begin{multline}
d(x, y) = D_1 \exp\left(-\frac{(x - x^1_1)^2}{2\sigma^2}\right) \exp\left(-\frac{(y - y^1_1)^2}{2\sigma^2}\right) + \\ 
D_2 \exp\left(-\frac{(x - x^2_1)^2}{2\sigma^2}\right)  \exp\left(-\frac{(y - y^2_2)^2}{2\sigma^2}\right)
\end{multline}
Notice, writing the distribution of the pair in such form, \textbf{we gain the knowledge about correlations, but we loose identities of the particles} since it is impossible now to identify the distribution of a specific particle in the mixture. In fact, a distribution for a single particle can exist only if it doesn't interact with anything else and didn't interact before, because if it did, correlations became part of the multi-particle density.
Generally, the simplest way to account for correlations is to write the multi-particle state vector as:
\begin{equation}
\ket{D} = \sum_{i_1, i_2, i_3, ..., i_N} C_{i_1, i_2, ..., i_N} \ket{D_{i_1}}
\ket{D_{i_2}}...\ket{D_{i_N}}, \label{eq:interacting_particles}
\end{equation}
where all summation indices are different. Now vectors $\ket{D}$ form a linear space - a direct product of vectors spaces corresponding to each particle. Please note, correlation coefficients $C_{i_1, i_2, ..., i_N}$ can change in time when a system evolves.

This situation appears only in QM since in CM we precisely know the positions and momenta of the particles and hence we should not worry about correlations.
\item The notion of interacting particles as it is stated above does not have any practical value because in order to trace the evolution of a system, we need to consider the time evolution of the whole Universe - no weak interactions can be ignored. For the practical purposes, we need to to be able to perform some sort of perturbation theory for weak interactions and maintain densities for our system. There are cases when we can do it by introducing potentials acting on a system. However, we can separate the distributions of the system and the rest of the Universe only under specific conditions. To be able to find these conditions, we need to investigate how a closed system evolves in time. We will do it in Section \ref{sec:weakly_systems}
\end{itemize}
\item What if an observer is not able to distinguish the particles? 
The density of two-particle state is constrained by: $\rho(\vec{x}_1, \vec{x}_2) = \rho(\vec{x}_2, \vec{x}_1)$.
This in turn imposes the constraint on the distribution: $d(\vec{x}_1, \vec{x}_2) =\pm d(\vec{x}_2, \vec{x}_1)$. Then $C_{12}=\pm C_{21}$ and equation \ref{eq:interacting_particles} reads:
\begin{equation}
\ket{D} = \frac{1}{\sqrt{2}}(\ket{D_1}\ket{D_2}\pm \ket{D_2}\ket{D_1})\label{eq:distribution_identical}
\end{equation}
or in coordinate representation:
\begin{equation}
d(\vec{x}_1, \vec{x}_2) = \frac{1}{\sqrt{2}}(d_1(\vec{x}_1) d_2(\vec{x}_2) \pm d_2(\vec{x}_2) d_1(\vec{x}_1)), \label{eq:distribution_identical}
\end{equation}
where $d_1(\vec{x})$ and $d_2(\vec{x})$ are distributions for the first and second particle respectively,
$\vec{x}_1$ and $\vec{x}_2$ are coordinates in orthogonal complements and the sign between two terms in equation \ref{eq:distribution_identical} can be plus or minus. 
Several observations can be made here:
\begin{itemize}
\item The source of equation \ref{eq:distribution_identical} is \textbf{invariance of the observable density}, not distribution, since the latter is unphysical.
\item In CM, the densities are localized\footnote{for the density centered at some point $\vec{x}_0$, we say that the corresponding particle is not smeared but rather located at the point $\vec{x}_0$}. Then:
\begin{equation}
\rho(\vec{x}_1, \vec{x}_2) = \frac{1}{2}(\delta(\vec{x_1}-\vec{x}^0_1) + \delta(\vec{x_2}-\vec{x}_2^0))
\end{equation}
In QM equation \ref{eq:distribution_identical} can lead to interference of two terms when squared but in CM it won't because the densities are orthogonal to each other.
\item In future, an observer might find a way to distinguish particles. Then, he should use the full distribution. By examining if equation \ref{eq:distribution_identical} works, an observer can test if the particles under investigation are identical. It would be interesting e.g. to derive identity of protons starting from quarks.  
\item For the identical particles, the sign in equation \ref{eq:distribution_identical} is meaningful since the density gets another term that depends on the product of the distributions of two particles and not on their densities. 
\begin{equation}
\rho(\vec{x}_1, \vec{x_2}) = \rho_1(\vec{x}_1) \rho_2(\vec{x}_2) + \rho_2(\vec{x}_1) \rho_1(\vec{x}_2) \pm 2 d_1(\vec{x}_1)d_2(\vec{x}_2) d_2(\vec{x}_1)d_1(\vec{x}_2)
\end{equation}
For $N$ particles, we require that the $N$ particle density doesn't change under any permutation and arrive at the Slater determinant \cite{bib:Slater}. 
\label{item:weaklyInteractingParticles}
\end{itemize}
\end{enumerate}

\section{Measurement}
\label{sec:measurement}
Suppose we perform a measurement of a system - an interaction of our test system with a probe system. This means that we take two non-interacting sub-systems (test and probe) and start turning on interactions. 
\begin{itemize}
\item In CM, initially the composite system is described by two vectors - one for the probe and one for the test system. After interaction, the probe doesn't change the test system - it just measures it's state.

\item In QM, initially, the composite system is described by the collection of two vectors (corresponding to two systems), but eventually the vector evolves into a vector in the direct product space. 

Since in CM, test systems don't change after interaction, let us try to keep this constraint and consider the set of vectors of the test systems that are multiplied by a real number after the interaction\footnote{The corresponding normalized vector characterizing the state of the system does not change.} \cite{bib:Zurek}. We can show that these vectors are orthogonal. For two such vectors $u_i$ and $u_j$ where $i\neq j$:
\begin{equation}
 \bra{u_i}\ket{u_j} = \bra{u'_i}\ket{u'_j} = \alpha_i \alpha_j \bra{u_i}\ket{u_j},
 \end{equation}
 where $u'_i$ and $u'_j$ are the vectors after interaction, $\alpha_i$ and $\alpha_j$ are some real numbers. Here we used the fact that the scalar product is preserved under orthogonal transformations.
\begin{equation}
\bra{u_i}\ket{u_j}(1-\alpha_i\alpha_j)=0.
\end{equation}
Since $\alpha_i$ and $\alpha_j$ are arbitrary and there is no reason to expect their inner product to be 1,
\begin{equation}
\bra{u_i}\ket{u_j} = 0 \label{eq:orthogonality}
\end{equation}
We can also define a linear orthogonal operator ($\hat{O}$) for which these vectors are eigenvectors.  We call this operator - a measurement operator. This is the same operator that is responsible for evolution of individual vectors. 

\end{itemize}

Vectors $u_i$ are delta functions (or delta symbols) in the eigenbasis of the measurement operator. Actually, they have some width related to the precision of the measuring apparatus - there is always an error associated to an apparatus. We can assume the distributions to be narrow Gaussians $\exp(-\frac{(u_i - u_i^0)^2)}{\sigma_0^2})$, where $\sigma_0$ is much less than characteristic values for $\min(\abs{u_i})$ of the measuring apparatus. This means that their eigenvalues correspond to the centers of the Gaussians, and that's why they provide the \textbf{measured} values of the corresponding variables. This fact is stated in QM books but only now its physical sense becomes clear.  

Generally speaking, eigenvectors of $\hat{O}$ are complex valued. In section \ref{sec:basics} we started by requiring our $d(\vec{x})$ to be real, however now we see that:
\begin{itemize}
\item We either need to restrict the set of our measurement operators and coordinates, so that the eigenvectors are always real.
\item Or we need to abandon the requirement that $d(\vec{x})$ is real.
\end{itemize}
It is not easy to satisfy the first requirement but the second requirement, although seems problematic from the physics perspective, actually is not. Indeed, as we found above, the only thing we measure are eigenvalues, and these are real. So we can start by not imposing any extra requirements and allowing the $d(\vec{x})$ to be complex. We will use $\psi(\vec{x})$ for what we called distribution function before and call it 'the wave function'. Also, since our wave function is complex, there is no sense of keeping the measurement operator orthogonal. We will use unitary operators for the measurement operators instead of orthogonal operators. Now, a particle is described by a vector (we also call it a state vector) in a complex Hilbert space. The norm of this vector is 1. The most obvious choice for the above density is $|\psi(\vec{x})|^2$. All earlier discussions concerning orthogonal operators stay intact when we move from orthogonal to unitary operators.

For $N$ identical interacting particles, we should make sure that any permutation does not change the density. It is easy to see that this can be accomplished only by the Slater determinant \cite{bib:Slater}. 

Let us expand the state vector. We can choose bases functions for example as delta functions $\delta(\vec{x} - \vec{x}_0)$. However, then the norm of the density will equal infinity (integral of square of a delta function). As it is stated above, the wave functions corresponding to the measurement operators ((eigenfunctions of its operator)) are narrow Gaussians, and despite the fact that they strictly speaking don't form an orthogonal basis, we can consider them 'almost' orthogonal. This may lead to some minor corrections in future and we should pay attention to them. Such a wave function for an apparatus measuring coordinate is then given by:
\begin{equation}
\psi(\vec{x}, \vec{x}^0) = \psi(\vec{x})_{\vec{x}_i^0} = (\frac{1}{\sigma_0\sqrt{2\pi}})^3 \exp{-\frac{(\vec{x} - \vec{x}^0)^2}{2 \sigma_0^2}}\label{eq:basis},
\end{equation}
and an arbitrary wave function can be written as:
\begin{equation}
\psi(\vec{x}) = \int C_{\vec{x}_0} \psi(\vec{x})_{\vec{x}_i^0} d\vec{x}_0,
\end{equation}
We call the corresponding coefficients of expansion - coordinate representation of the state vector.
There is also infinitely many possible bases, each yields a description of our system and may be associated with a set of eigenfunctions of some measurement apparatus. One of such basis is a set of harmonic functions:
\begin{equation}
\psi_k(x) = C \exp(i\vec{k}\vec{x} + \phi), \label{eq:Furie}
\end{equation}
where C is a normalization constant. The phase $\phi$ is just an overall constant factor and can be absorbed into coefficient $C$. However, these functions are not normalizable in the same wave as delta functions are not normalizable. Therefore, let us multiple each such function by a wide Gaussian which tends to zero at large\footnote{Much larger than the values of k for base functions that contribute to expansion} values of $k$.
\begin{equation}
\psi_{\vec{k}}(\vec{x}) \sim \exp(-i\vec{k}\vec{x}) \exp{-\frac{\vec{k}^2}{2 \sigma_{k_0}^2}}, \label{eq:Furie}
\end{equation}
Let us introduce a new coordinate $p$ and define it:
\begin{equation}
\vec{p} = \hbar\vec{k}
\end{equation}
where $\hbar$ is an arbitrary number called "Planck's constant". This transformation is just a scale to use more convenient units for measuring $\vec{k}$.  Defining $\sigma_{k_0} = 1/\sigma_0$, transition from coordinate representation to a new representation is then given by: 
\begin{equation}
\psi_{\vec{p}} (\vec{x})\sim \int\exp(-i\frac{\vec{p}\vec{x}}{\hbar})\psi(\vec{x}, \vec{x}'^0)d^3 x'^0 \label{eq:Furie}
\end{equation}
If we define an operator $\vec{p}$ such that 
\begin{equation}
\hat{\vec{p}} = -i\hbar \nabla, \label{eq:MomentumOperator}
\end{equation}
we get using equation \ref{eq:Furie}:
\begin{equation}
\psi(\vec{p}) = \hat{p} \psi(\vec{x}) \label{eq:MomentumOperatorAction}
\end{equation}

At this point, we suspect that $\vec{p}$ is related to one of our fundamental variables in QM with Gaussian factor related to the resolution of an apparatus measuring this variable\footnote{Because equation \ref{eq:MomentumOperatorAction} describes an apparatus that measures some variable $\vec{p}$ transforming wave function $\psi(\vec{x})$ before measurement to $\psi(\vec{p})$ after the measurement.}. The closest analogy in CM is wave vector or momentum \footnote{According to equation \ref{eq:Furie}, $\vec{k}$ has dimension $1/m$. If we want it to be momentum, our multiplicative parameter should have dimension $\frac{1/m}{m/s} = \frac{s}{m^2}$.}.  By matching QM dynamics to CM, we will be able to see in section \ref{sec:Mapping} that this is indeed the case. 

Here is the summary of what we achieved so far.
\begin{enumerate}
\item We started by introducing a function $\rho(\vec{x})$ corresponding to the density of a particle, such that $\int \rho(\vec{x}) d^3 x = 1$.
\item This allowed us to represent a particle as a vector in Hilbert space. We then concluded that in order for the number of particles to be invariant, we need to change our definition to that in equation \ref{eq:distribution}.
\item In this section we realized that the $d(\vec{x})$ can also be complexed value and called it $\psi(\vec{x})$. Now $\psi(\vec{x})$ does not have a direct physical meaning, but $|\psi(\vec{x})|^2$ corresponds to the density we started from.
\item We then stated that there are infinitely many other bases, and state vectors can be expanded to any of them. We call such expansions - representations and gave an example of a momentum representation.
\end{enumerate}

\section{Time evolution operator}
\label{sec:evolution_cont}
Let us get back to the evolution operator and see if we can figure out its exact form.
\begin{itemize}
\item Since now we have a complex-valued function, our rotations will be represented by unitary transformations : $\psi_k \to \psi_k(t) = \sum_j\exp(i A_{kj})\psi_j$,  where $A_{kj}$ is a hermitian matrix.
\item Let us diagonalize this matrix. We will get $A\to I * \lambda$, where I is a identity matrix, and $\lambda$ is the vector of eigenvalues of the operator $A$. They can be interpreted as angles of rotations around corresponding axes. The simplest non-trivial dependence of such rotation on time is rotation with constant angular velocities. Therefore we state that $\lambda_k = \omega_k * t$, where $\omega_k$ is a constant. The evolution in such a basis will be given by:
\begin{equation}
\psi'_k(t) = \psi'_k(0) \exp(-i \omega_k t), \label{eq:evolutiondiag}
\end{equation}
where $\psi'_k(t)$ is the $k$'s component of $\psi'(t)$ in the system of coordinates where $A$ is diagonal. 
It is clear from above equation that the state vector 'turns' the $k$'s coefficient $c_k$ with constant angular velocity $\omega_k$. 
\item The evolution of the state vector then takes the form :
\begin{equation}
\psi(t) = \psi(0) \exp(-i \hat{A} t) \label{eq:evolution}
\end{equation}
For $-\hat{A}\hbar$ being Hamiltonian, we get the Schr\"{o}dinger equation.
\end{itemize}

\section{Maximum entanglement and decoherence}
\label{sec:decoherence}

Let us come back to interactions of the two subsystems - test and probe:
\begin{equation}
\ket{\psi} = \sum_{i} C_i \ket{u_i} \ket{\chi}  \label{eq:systemBefore}
\end{equation}
The test subsystem is expanded to the basis of the measurement operator, and $\chi$ is the wave function of the probe subsystem before the measurement.
After the measurement we have:
\begin{equation}
\ket{\psi'} = \sum_{i} D_i \ket{u_i} \ket{v_i},   \label{eq:systemAfterMeas}
\end{equation}
where $D_i$ are complex value coefficients, $u_i$ are the test eigenfunctions and $v_i$ are some vectors in the second subsystem. Note that $u_i$ didn't change. For an arbitrary interaction we can assume that $v_i$ are random normalized vectors in Hilbert space. Let us calculate $\bra{v_i}\ket{v_j}$ where $i\neq j$.
For two arbitrary normalized vectors in a $N$ dimensional space, we can think of their inner product as a projection of $v_j$ on $v_i$. For a random vector $v_j$, we assume that absolute values of all projections are equal on average. Thus,
$v_1 = v_2 = ... = v_n$ and hence since $\sum_i v_i^2 = 1$, on average: 
\begin{equation}
\bra{v_i}\ket{v_j} = 1 / \sqrt{N}. \label{eq:scalarProduct}
\end{equation}
For $N >> 1$, 
\begin{equation}
\bra{v_i}\ket{v_j} \sim 0. \label{eq:orthogonality}
\end{equation}
For the second subsystem with many degrees of freedom, we conclude that the inner product is zero. 
When $v_j$ are almost orthogonal, we say that the system decohered, and we call the whole process - decoherence \cite{bib:Zurek}. Equation \ref{eq:systemAfterMeas} with orthonormal $u_i$ and $v_j$ are called Schmidt decomposition \cite{bib:Shmidt}. 
The time it takes a system to decohere depends on the number of degrees of freedom and mass of the subsystems, and generally is very small (below what we can measure) for large systems (see equation \ref{eq:scalarProduct}). 

For large probes, $\ket{v_i}$ in equation \ref{eq:systemAfterMeas} are orthogonal. Since both $\ket{u_i}$ and $\ket{v_i}$ form an orthogonal set, each term $D_i \ket{u_i} \ket{v_i}$ is completely independent of all others. Therefore, after the interaction, the system consists of $N$ completely independent terms - we will call them 'branches'.

One immediate consequence is that once two subsystems interacted, they are always mixed, we won't be able to find individual subsystems in the mixture. It might mean that there are no unmixed systems in the world and everything is entangled with everything. The wave function of the Universe then can be represented by:
\begin{equation}
\psi_{universe} = \sum_{branches_i} \prod_{objects_j} \ket{v_{branch_i, object_j}}, \label{eq:Universe}
\end{equation}
where $\bra{\prod_{objects_j}v_{branch_{i1}, object_j}}\ket{\prod_{objects_{j'}}v_{branch_{i2}, object_{j'}}} = 0$ for objects that decohered. 

\section{Measurement and the Quantum Postulate}
\label{sec:measurement}
We notice an important consequence of the above discussion : unlike in CM where we can always outline subsystems in the composite system, in QM the subsystems are mixed. Moreover, according to the equation \ref{eq:systemAfterMeas}, each system having interacted with another large system, gets entangled. From this point in time, the state vectors of individual systems don't exist anymore even when the systems are far from each other and not interacting. 

Let us define an apparatus as a quantum system which decoheres fast\footnote{Decoherence time is much smaller than effective interaction time}. Usually apparatuses are large and have many degrees of freedom. After a particle interacts with an apparatus, the wave function of the apparatus-particle composite system is given by equations \ref{eq:systemAfterMeas} and \ref{eq:orthogonality}. There are multiple completely independent branches in this decomposition. 

Let us take a concrete example : a double slit experiment followed by two films parallel and very close to each other\footnote{In order for the electron wave functions not to evolve considerably while electron travels between the films}. Let's assume that an electron penetrates the first film leaving a spot on the film, and then gets absorbed by the second film. There is also an observer (who has many degrees of freedom), observing the spots on two films. 

Wave function of the system after an electron gets absorbed by the second film looks like.
\begin{equation}
\bra{\vec{x}}\ket{system} = \bra{\vec{x}}(\ket{ele} \ket{film_1} \ket{film_2} \ket{\gamma} \ket{obs}), \label{eq:measurement}
\end{equation}
where $\ket{\gamma}$ is a photon state vector that delivers information to observer's eyes\footnote{Since photons are relativistic objects and their number does not conserve, the wave functions of photons are strictly speaking not defined}, and $\vec{x}$ is a coordinate along film's axes. Also, $\bra{x}\ket{ele}$, $\bra{x}\ket{film_1}$, $\bra{x}\ket{film_2}$ are Gaussians centered at coordinate $\vec{x}$. Please note, the observer is no longer decoupled from the apparatus, moreover, it is entangled with the $e-film_1-film_2-\gamma$ system. Therefore, it is wrong to imagine an observer looking at the system because the observer and the system don't exist anymore, but the entangled system does. 

We should read equation \ref{eq:measurement} as continuum of observers $obs_{\vec{x}}$ each looking at the corresponding  $\bra{x}\ket{ele}$, $\bra{x}\ket{film_1}$, $\bra{x}\ket{film_2}$. Each observer with index $\vec{x}$ will see a spot on the first film at point $\vec{x}$ and on the second film at the same point $\vec{x}$. Also if an observer $\vec{x}$ sees a spot at $\vec{x}$ and sees another observer, then another observer also sees the spot at the same point $\vec{x}$. In other words, there is no collapse, wave function evolves according to Schr\"{o}dinger equation, however the appearance of collapse is achieved due to entanglement of the observer with the system. 

Let us take another example : EPR experiment \cite{bib:EPR}. In this experiment, the pair of particles are not only entangled with each other but are also entangled with the observers. Therefore, one pair of observers would see the first branch, and another pair would see another branch. No superluminal velocities are involved since the branches existed even before the measurement.

\section{Born's rule}
In section \ref{sec:measurement} we realized that a particular observer can't see the whole wave function of a subsystem simply because this wave function doesn't exist for the subsystem, instead the wave function of the whole system exists. However, an observer attached to a particular branch, can see a system corresponding to this branch. 

Now the natural question arises: "What is the probability that a particular observer finds himself in a branch with number $i$ observing a subsystem $i$?" This question cannot be addressed by Schr\"{o}dinger equation alone since this equation gives the time evolution of the wave function of the whole system, and doesn't provide any information about probabilities\footnote{Indeed, so far we found out that wave function is related to the density but never discussed its physical meaning. Schr\"{o}dinger equation is just an equation that takes the wave function at time zero, and returns the wave function at time t}. At this point, we need to return to our original notion of density. We introduced $\rho(\vec{x})$ as a density of our electron smeared all over the space. Since then, there were two significant changes:
\begin{enumerate}
\item We introduced a complex valued wave function
\item This wave function exists only for non-interacting systems which were never entangled, and it is wrong to assign wave function to a subsystem after interaction.
\end{enumerate}
The second change forces us to seek for the probabilities corresponding to a wave function of the system which was never entangled. But this function is complexed valued. 

Let us consider $M$ sub-systems after they all decohered. Suppose we have $N$ branches in the wave function of the composite system::
\begin{equation}
|\psi_{system}> = \sum_{i = 1 ... N} \rho_i \exp(i\phi_i) |\psi^1_i> |\psi^2_i> ... |\psi^M_i>, \label{eq:branches}
\end{equation}
where $\rho_i$ is the absolute value of the $i$ complex coefficient and $\phi_i$ is its phase. Also, $\phi_i$ is the sum of all phases for all sub-systems. 

Let us recall how we started. Our $\psi(\vec{x})$ came from the distribution $d(\vec{x})$ whose square defines a density in a point $\vec{x}$. We can just state that our system is smeared over the branches - each branch gets a weight equal to the corresponding density. The density for the branch is then: $\rho_i(\vec{x}) = |\psi_i(\vec{x})|^2$. We can then assume that there are multiple copies of the same observer attached to a branch due to degeneracy\footnote{In equation \ref{eq:Universe}, degeneracy of a branch is defined as the number of terms in sum corresponding to the same observer.}, and their number is proportional to this density. Therefore, the density equals probability for this observer to find himself in the branch.

We also need to prove that our definition of the probabilities does not create contradictions down the line. In \cite{bib:Zurek}, a derivation of Born's postulate from symmetry consideration is provided which can also serve as such a prove.

\section{Mapping to CM and exact form of the evolution operator.}
\label{sec:Mapping}
In CM, we use equations of motion to trace the evolution of the system. In QM, as we found out in section \ref{sec:evolution}, we use the time evolution operator. It is important to obtain CM starting from QM and taking some limit. We will define the limit as follows: suppose we measure some property of a system multiple times\footnote{Of course, after the measurement, we create a mixed state of the system with the apparatus so it is practically impossible to get back to the initial state vector. So in practice we prepare many systems with identical state vector and let them interact with the set of identical apparatuses.}. Each time after a measurement, we get some measured value. Averaging these values will give us some value. For a classical system, this value doesn't change and equals the measured value.\footnote{Actually, there will be some scatter due to the imperfect apparatus, but this scatter is not related to the quantum mechanical scatter.} Therefore, QM averages and equations of motion coincide with classical values for classical systems and equations of motion.

According to the previous section, each time measuring the system we just get its state vector corresponding to one of the branches. When we average the results, we give a weight to each branch, corresponding to the Born's rule \cite{bib:Zurek}. Thus, we arrive at the well known form for expected values which maps QM to CM.
\begin{equation}
M = \sum_{branch} \rho_{branch} M_{branch} = \bra{\psi}\hat{M}\ket{\psi}, \label{eq:expectation}
\end{equation}
where $\hat{M}$ in the $rhs$ of equation \ref{eq:expectation} is an operator of the apparatus that measures $M$ - its classical value.

For interacting subsystems, generally speaking we cannot separate each subsystem in the 
many-particle wave function. Therefore, when trying to match QM to CM, we should start from systems not entangled with anything and consider their evolution. Suppose we have two interacting particles, the classic potential between them is given by $U(x_1, x_2)$.

We know that the classical energy, total classical momentum and total classical angular momentum for a closed system conserve:
\begin{eqnarray}
\frac{dE}{dt} = 0 \\
\frac{d\vec{p}}{dt}= 0 \\
\frac{d\vec{L}}{dt}= 0
\end{eqnarray}
Let us use equation \ref{eq:evolution} to find the evolution of the expectation value for the variable $M(t)$:
\begin{multline}
\frac{dM(t)}{dt} = \frac{d\bra{\psi(\vec{x}_1, \vec{x}_2, t)}\hat{M}\ket{\psi(\vec{x}_1, \vec{x}_2, t)}}{dt} = \\ \frac{d\bra{\psi(x_1, x_2, 0)\exp(i\hat{A}t)}\hat{M}\ket{\exp(-i\hat{A}t)\psi(0)}}{dt} = 
\\ \bra{i\hat{A}\psi(\vec{x}_1, \vec{x}_2,0)\exp(i\hat{A}t)}\hat{M}\ket{\exp(-i\hat{A}t)\psi(\vec{x}_1, \vec{x}_2, 0)} + \\ \bra{\psi(\vec{x}_1, \vec{x}_2, 0)\exp(i\hat{A}t)}\hat{M}\ket{\exp(-i\hat{A}t)\psi(\vec{x}_1, \vec{x}_2, 0)(-i\hat{A})} = \\
-i\bra{\psi(\vec{x}_1, \vec{x}_2, t)}[\hat{M}A]\ket{\psi(\vec{x}_1, \vec{x}_2, t)},
\end{multline}
This means that if we choose:
\begin{equation}
\hat{A} = f(\hat{H}),
\end{equation}
or
\begin{equation}
\hat{A} = f(\hat{\vec{p}}),
\end{equation}
or
\begin{equation}
\hat{A} = f(\hat{\vec{L}}),
\end{equation}
where $\hat{H}$ is the operator corresponding to energy, we will restore conservation of energy, momentum or angular momentum in classical limit. $\hat{A} = f(\hat{\vec{p}})$ doesn't depend on coordinates, so a momentum of each individual interacting particle will conserve which is clearly not what we expect. Using Occam's razor \cite{bib:Occam}, let us choose $\hat{A} = f(\hat{H}) = C \hat{H}$, where C is some constant:
\begin{equation}
\hat{A} = C \hat{H} = C (\frac{\hat{\vec{p}}_1^2}{2 m} + \frac{\hat{\vec{p}}_2^2}{2 m_2}+ U(\hat{\vec{x}}_1, \hat{\vec{x}}_2)) = 
C(-\frac{\hbar^2}{2 m_1} \Delta -\frac{\hbar^2}{2 m_2} \Delta+ U(\hat{\vec{x}}_1, \hat{\vec{x}}_2)), \label{eq:Hamiltonian}
\end{equation}
and check if the total momentum and angular momentum also conserve. 
We can use our new form for the time evolution operator to recover the Ehrenfest equations \cite{bib:Ehrenfest} for values of momentum $p$ and coordinate $x$ of the test particle\footnote{The individual wave functions don't exist, but we can still find the average properties for each particle by averaging the corresponding operators using two-particle wave function}. 
\begin{eqnarray}
\frac{d\vec{p}_1}{dt} = -\nabla_1U(\vec{x}_1, \vec{x}_2), \label{eq:Ehrenfest1} \\
\frac{d\vec{p}_2}{dt} = -\nabla_2U(\vec{x}_2, \vec{x}_1), \label{eq:Ehrenfest2}
\end{eqnarray}
\begin{eqnarray}
\frac{d\vec{x}_1}{dt} = \frac{\vec{p}_1}{m} \label{eq:Ehrenfest3} \\
\frac{d\vec{x}_2}{dt} = \frac{\vec{p}_2}{m} \label{eq:Ehrenfest4}
\end{eqnarray}
with $C=\frac{1}{\hbar}$.
Therefore, equations \ref{eq:evolution} now reads:
\begin{equation}
\psi(t) = \psi(0) \exp(-i \frac{\hat{H}}{\hbar} t), \label{eq:evolution_improved}
\end{equation}
and from here we obtain  the Schr\"{o}dinger equation:
\begin{equation}
i\hbar \frac{\partial \psi}{\partial t} = \hat{H}\psi.
\end{equation}
The conservation of total momentum $p_1+p_2$ follows from CM\footnote{Because we obtained CM equations of motions and the conservation laws are just the consequence}. 

It is worthwhile to notice that in CM, Planck's constant disappears, it cancels out. However, an exact value of $\hbar$ in QM actually defines our time evolution in \ref{eq:Hamiltonian} and \ref{eq:MomentumOperator}.

In many textbooks, Schr\"{o}dinger equation is postulated making a reader wonder why Schr\"{o}dinger equation takes this and not any other form. We tried to arrive at the Schr\"{o}dinger equation naturally and prove its consistency with CM.

\section{Weakly interacting systems} \label{sec:weakly_systems}
In Section \ref{sec:multiple_particles}, we were trying to find such weakly interacting subsystems of the bigger subsystems, for which each of the subsystems could be identified as a separate subsystem subject to Schr\"{o}dinger equation with some potential. 

However, equation \ref{eq:Universe} tells us that everything is entangled with everything, and there is no way to factor our single particles from the wave function of the Universe. How to resolve this contradiction?

Equation \ref{eq:measurement} implies that the wave function consists of orthogonal branches, and each branch has a set of observers attached to it. These observes see only this branch and for them, the particles are not entangled - the individual particles can be identified.

After some time however, if the particles interact, the observers will not be able to find individual particles in the corresponding branch. So the above question about weakly interacting subsystems should be posed for one branch. Initially a branch consists of individual particles. Under which conditions after time $t$, the particles can still be identified? If these conditions are not satisfied, the branch will again branch out to sub-branches - each sub-branch will have a new set of observers attached to it.

There are several important conditions for preventing sub-branching:
\begin{enumerate}
\item Let's consider two interacting particles. Their interaction is described by a potential $U(\vec{x}_1, x_2)$, where $\vec{x}_1$ is the coordinate of the first particle, and $\vec{x}_2$ is the coordinate of the second particle.
The Schr\"{o}dinger equation can be written as follows:
\begin{equation}
i\hbar \frac{\partial \psi(\vec{x}_1, \vec{x}_2, t)}{\partial t} = (\frac{\bar{\vec{p}}_1^2}{2 m_1} + \frac{\bar{\vec{p}}_2^2}{2 m_2} + U(\vec{x}_1, \vec{x}_2)) \psi(\vec{x}_1, \vec{x}_2, t)
\end{equation}
For the linear potentials of the form $U(\vec{x}_1, \vec{x}_2) = \alpha |\vec{x}_1 - \vec{x}_2|$, the Schr\"{o}dinger equation can be factorized:
\begin{eqnarray}
i\hbar \frac{\partial\psi(\vec{x}_1, t)}{\partial t} = (\frac{\bar{\vec{p}}_1^2}{2 m_1} + \alpha \vec{x}_1) \psi(\vec{x}_1, t) \\
i\hbar \frac{\partial\psi(\vec{x}_2, t)}{\partial t} = (\frac{\bar{\vec{p}}_2^2}{2 m_1} - \alpha \vec{x}_2) \psi(\vec{x}_2, t),
\end{eqnarray}
where $\psi(\vec{x}_1, \vec{x}_2, t) = \psi(\vec{x}_1, t) \psi(\vec{x}_2, t)$.
In this case (e.g. a particle in an electrostatic field), the particle retains it's identity, but Schr\"{o}dinger equation includes an external potential. 

\item If a potential changes slowly, we can use only the first term in Taylor's expansion and approximately factor out multi-particle wave function. 
For a potential to change slowly, the second term in the Taylor's expansion of the potential should be much smaller than the the first term. In other words, for two particles, from Schr\"{o}dinger equation:
\newcommand\at[2]{\left.#1\right|_{#2}}
\begin{equation}
\frac{1}{2} |\int \at{\frac{\partial^2 U} {\partial r^2}}{r^0} (r-r_0)^2 \psi(r) dr | \ll |\int \at{\frac{\partial U}{ \partial r}}{r_0} (r-r_0) \psi(r) dr |, \label{eq:slow_changing_potential}
\end{equation}
where $r_0$ is the distance between the particles under consideration.
Suppose, $U(r) = -\frac{\alpha}{r}$, where $\alpha > 0$. Also suppose that the wave function of the test particle is a Gaussian with width $\sigma$, corresponding to an electron, and the wave function of the probe is a narrow Gaussian, corresponding to an atom (Figure \ref{fig:TestProbe}).
\begin{figure}
\raggedright
\includegraphics[scale=0.5]{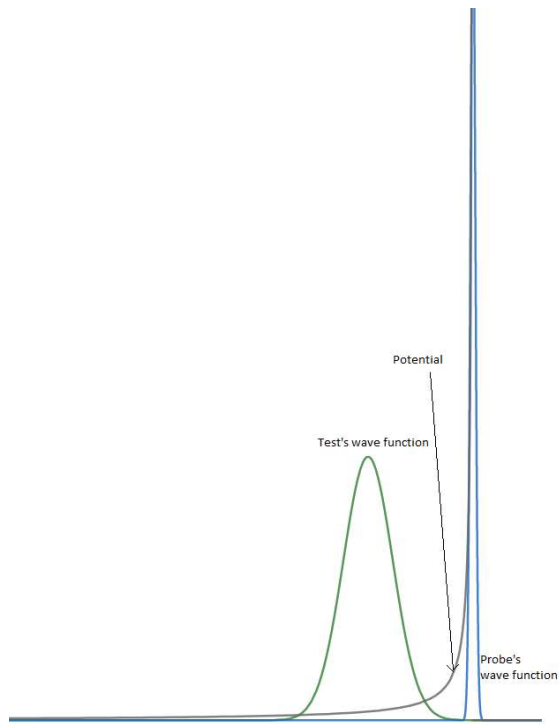}
\caption{Two interacting particles. The distance between the particles is chosen in such a way that the potential is almost linear. $\omega = \frac{1}{3}$.}
\label{fig:TestProbe}
\end{figure}

Therefore, equation \ref{eq:slow_changing_potential} reduces (up to coefficients) to:
\begin{equation}
\frac{\alpha}{r_0^3} \sigma^2 << \frac{\alpha}{r_0^2} \sigma
\end{equation}
or
\begin{equation}
\sigma \ll r_0
\end{equation}
To be conservative, we can introduce a coefficient $\omega < 1$ representing the degree of linearity of our potential and write the above equation as:
\begin{equation}
\sigma = \omega r_0, \label{eq:powerpotential}
\end{equation}

The important conclusion we can draw from equation \ref{eq:powerpotential} is that \textbf{we can talk about individual particles only when the widths of their individual wave functions are much smaller than the distance between them}.

\item For a particle in the constant magnetic field, only kinetic terms change, and the system can still be factorized. 

\item Wave functions of classical objects are narrow. This means that positions and momenta of each object is set and there is no place for correlations, hence multiobject wave function can be factorized to yield a product of individual wave functions. Therefore, each classical object is a subject of classical equations of motions (see also Section \ref{sec:ev_classical}).
\end{enumerate}

However, generally an interacting particle cannot be separated from the environment. Say, for two particles wave function, we can change the variables : $\vec{x}_1, \vec{x}_2 \to \vec{r} = \vec{x}_1 - \vec{x}_2$. For the new variable, Schr\"{o}dinger equations reads:
\begin{equation}
i\hbar \frac{\partial\psi(\vec{r}, t)}{\partial t} = (\frac{\bar{\vec{p}}^2}{2 \mu} + U(\vec{r})) \psi(\vec{r}, t),
\end{equation}
where $\mu = \frac{m_1 m_2}{m_1 + m_2}$ is a reduced mass. However, despite the Schr\"{o}dinger equation is now 1-dimensional, we didn't factorize the system, the particles are still entangled.

This very important fact is the source of the Quantum postulate and is overlooked in many QM discussions.

\section{Composite systems}
\label{sec:composite}
Suppose we have a system composed of many particles. How do we treat such a system? Should we consider equations of motion of each individual particle to figure out the behavior of the whole system?
\begin{itemize}
\item{In CM, we prove that Newtonian equations of motion apply to the center of mass of the system, and we can ignore the fact that the system is composite when considering its evolution\footnote{For figuring out internal time evolution of the system (e.g. rotations), we need to take into account its internal interactions, but to get the evolution of the system as a whole, we can ignore them.}.}
\item{In QM, this question needs some clarification because as stated above, we can talk about standalone particles only under specific conditions. Suppose we have a system composed of many particles. This system is far enough from all other particles, so for each particle inside this system, equation \ref{eq:powerpotential} holds. Can we treat this system as a particle, namely as having a wave function and evolving according to Schr\"{o}dinger equation with some potential?

The answer to this question usually quietly assumed to be true, but it needs to be proven. 
For a system of $N$ particles in a potential $U$, Schr\"{o}dinger equation reads:
\begin{equation}
i\hbar \frac{\partial\psi(\vec{r}_1,\vec{r}_2,...,\vec{r}_N , t)}{\partial t} =  (\sum _{i=1..N}\frac{\hat{\vec{p}_i}^2}{2 m_i} + U(\vec{r}_1,\vec{r}_2,...,\vec{r}_N)) \psi(\vec{r}_1,\vec{r}_2,...,\vec{r}_N , t), \label{eq:seforNParticles}
\end{equation}
where
\begin{equation}
U(\vec{r}_1,\vec{r}_2,...,\vec{r}_N) = \sum_{pairs(i,j) = 1..N} V^{int}(\vec{r}_i - \vec{r}_j) + \sum_{i=1..N} V^{ext}(\vec{r}_i), \label{eq:potential}
\end{equation}
and $V^{int}$, $V^{ext}$ are potentials corresponding to interactions between particles and interaction between the composite system and other particles respectively. Guided by CM solution, we want to factor out wave function of the center of mass of the system from the $N$-particle wave function:
\begin{equation}
\psi(\vec{r}_1,\vec{r}_2,...,\vec{r}_N , t) = \psi(\vec{r}_C, t) \psi(\vec{r}_1 - \vec{r}_C,\vec{r}_2 - \vec{r}_C,...,\vec{r}_N - \vec{r}_C , t), \label{eq:factorization}
\end{equation}
where:
\begin{equation}
r_c = \frac{\sum_i m_i \vec{r}_i}{\sum m_i}. \label{eq:rc}
\end{equation} 
Plugging equations \ref{eq:rc}, \ref{eq:factorization} and \ref{eq:potential} into equation \ref{eq:seforNParticles}, we get for the wave function of the center of mass:
\begin{multline}
i\hbar \frac{\partial\psi(\vec{r}_C, t)}{\partial t} \psi(\vec{r}_1 - \vec{r}_C,\vec{r}_2 - \vec{r}_C,...,\vec{r}_N - \vec{r}_C , t) + \\ i\hbar \frac{\partial\psi(\vec{r}_1 - \vec{r}_C,\vec{r}_2 - \vec{r}_C,...,\vec{r}_N - \vec{r}_C , t)}{\partial t} \psi(\vec{r}_C, t) = \\
-\frac{\hbar^2}{2 \sum m_i} \Delta \psi(\vec{r}_C, t) \psi(\vec{r}_1 - \vec{r}_C,\vec{r}_2 - \vec{r}_C,...,\vec{r}_N - \vec{r}_C , t) \\
- \sum_i  \frac{\hbar^2}{ m_i} (\nabla \psi(\vec{r}_C, t) \nabla \psi(\vec{r}_1 - \vec{r}_C,\vec{r}_2 - \vec{r}_C,...,\vec{r}_N - \vec{r}_C , t)) - \\
\sum_i \frac{\hbar^2}{2 m_i}  \psi(\vec{r}_C, t) \Delta_i \psi(\vec{r}_1 - \vec{r}_C,\vec{r}_2 - \vec{r}_C,...,\vec{r}_N - \vec{r}_C , t)
+ \\ U(\vec{r}_1,\vec{r}_2,...,\vec{r}_N) \psi(\vec{r}_C, t) \psi(\vec{r}_1 - \vec{r}_C,\vec{r}_2 - \vec{r}_C,...,\vec{r}_N - \vec{r}_C , t) = \\
-\frac{\hbar^2}{2 \sum m_i} \Delta \psi(\vec{r}_C, t)\psi(\vec{r}_1 - \vec{r}_C, \vec{r}_2 - \vec{r}_C,...,\vec{r}_N - \vec{r}_C , t) - \\
\sum_i \frac{\hbar^2}{2 m_i}  \psi(\vec{r}_C, t) \Delta_i \psi(\vec{r}_1 - \vec{r}_C,\vec{r}_2 - \vec{r}_C,...,\vec{r}_N - \vec{r}_C , t)
+ \\
U(\vec{r}_1,\vec{r}_2,...,\vec{r}_N) \psi(\vec{r}_C, t) \psi(\vec{r}_1 - \vec{r}_C,\vec{r}_2 - \vec{r}_C,...,\vec{r}_N - \vec{r}_C , t). \label{eq:composite_der}
\end{multline}
In derivation \ref{eq:composite_der}, the scalar product of two gradients vanishes because:
\begin{multline}
\sum_i \frac{\hbar^2}{m_i} (\nabla \psi(\vec{r}_C, t) \nabla \psi(\vec{r}_1 - \vec{r}_C,\vec{r}_2 - \vec{r}_C,...,\vec{r}_N - \vec{r}_C , t)) = \\
\sum_i \frac{\hbar^2}{m_i}\frac{m_i}{\sum_j m_j}\nabla_{\vec{r}_C}\psi\sum_k \nabla_{\vec{r}_k - \vec{r}_C}\psi(\delta_{ki} - \frac{m_i}{\sum_j m_j}) = \\
\frac{\hbar^2}{\sum_j m_j}\nabla_{\vec{r}_C}\psi\sum_k\nabla_{\vec{r}_k - \vec{r}_C}\psi\sum_i(\delta_{ki} - \frac{m_i}{\sum_j m_j}) = 0
\end{multline}
We can break equation \ref{eq:composite_der} into two parts:
\begin{equation}
i\hbar \frac{\partial\psi(\vec{r}_C, t)}{\partial t} = -\frac{\hbar^2}{2 \sum m_i} \Delta \psi(\vec{r}_C, t) + \sum_i V^{ext}(\vec{r_i})\psi(\vec{r}_C, t), \label{eq:composite}
\end{equation}
and
\begin{multline}
i\hbar \frac{\partial\psi(\vec{r}_1 - \vec{r}_C,\vec{r_2} - \vec{r}_C,...,\vec{r}_N - \vec{r}_C , t)}{\partial t} = \\
\sum_i \frac{\hbar^2}{2 m_i} \Delta_i \psi(\vec{r}_1 - \vec{r}_C,\vec{r}_3 - \vec{r}_C,...,\vec{r}_N - \vec{r}_C , t) + \\
\sum_{i, j = 1..N} V^{int}(\vec{r}_i - \vec{r}_j)\psi(\vec{r}_1 - \vec{r}_C,\vec{r}_2 - \vec{r}_C,...,\vec{r}_N - \vec{r}_C , t).
\end{multline}

It follows from equation \ref{eq:composite} that when considering evolution of composite systems, we can decouple the wave function of its center of mass from the wave function related to internal movements.
Therefore, we can ignore particles inside these systems and just solve Schr\"{o}dinger equation for their centers of mass. We need to point out that this is possible because the product of two gradients vanishes. If we were to choose a different Hamiltonian than \ref{eq:Hamiltonian}, we might not been able to get this important property of the composite system, and in order to understand evolution of the composite objects, we would need to solve equations of motion for elementary objects. Clearly, our world is modular - generally speaking, we don't need to talk about electrons when considering stones.}
\end{itemize}

\section{Interaction of two particles}
\begin{figure}
\centering
\includegraphics[scale=0.5]{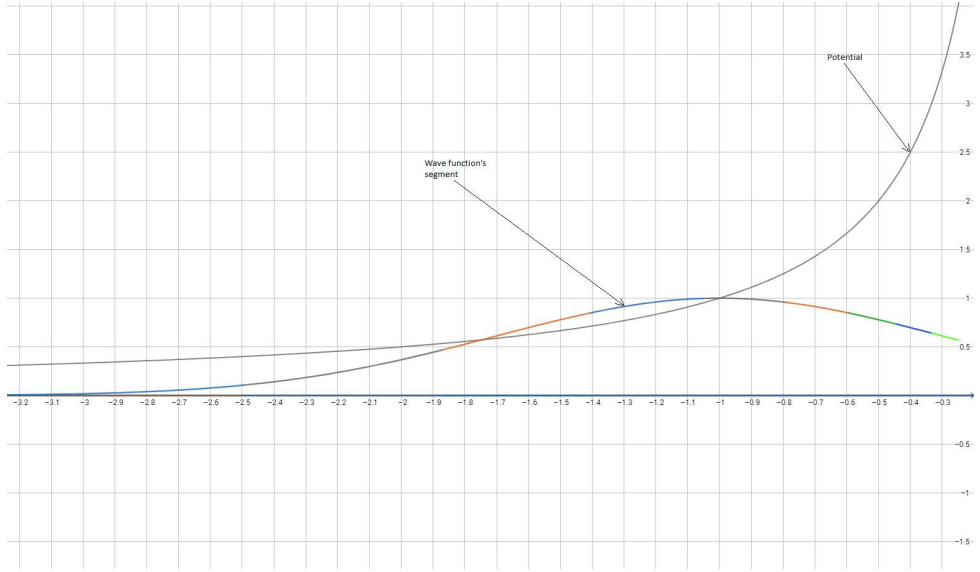}
\caption{The wave function of a test particle overlayed with the Coulomb potential between the particles. The wave function is split into a series of segments marked with different colors. The potential is approximately linear for the range of distances spanned by each segment. The units are arbitrary and numbers on the axes are presented only for better reading. The distance between two consecutive segments is chosen in such a way that the potential is almost linear $\omega = \frac{1}{3}$.}
\label{fig:Train}
\end{figure}

Let us consider two particles with masses $m_{probe}$ and $m_{test}$ running into each other, where the wave function of the probe is much narrower than the distance between the particles. The probe particle will represent an apparatus measuring some properties of the test particle.
When the particles are far from each other so that equation \ref{eq:powerpotential} holds, they can be separated, and the evolution of each particles is described by Schr\"{o}dinger equation. When the particles are close so that the width of their wave functions are larger than the distance between them, this approximation is invalid. Solving Schr\"{o}dinger equation for the pair is possible. However, if we want to represent this interaction as a measuring process, we want to be able to factor out individual particles after this interaction. To achieve this, we can split the test's wave function in multiple segments so that the potential for each such segment is almost linear, and, according to equation \ref{eq:slow_changing_potential}, we will be able to factor out individual particles from the multi-particle wave function. We say that the wave function of the test particle is a superposition of wave functions corresponding to individual segments\footnote{Each wave function is defined for the same range of coordinates as the corresponding segment.} (Figure \ref{fig:Train}). 
\begin{equation}
\psi^{test} = \sum_i C^{test}_i \psi^{test}_i
\end{equation}
The wave function of the system before interaction\footnote{Assuming these particles never interacted before} then reads:
\begin{equation}
\psi^{system} = \psi^{probe} \psi^{probe} = \sum_i (C^{test}_i \psi^{test}_i) \psi^{probe} = \sum_i (C^{test}_i \psi^{test}_i \psi_i^{probe}) \label{eq:train}
\end{equation}
where $\psi^{test}_i$ are chosen in such a way that for each $\psi^{test}_i$ the potential is approximately linear (equation \ref{eq:slow_changing_potential}), and at time $t=0$, $\psi_i^{probe} = \psi^{probe}$ for all values of $i$.
Thus, Schr\"{o}dinger equation for two-particle wave function can be approximated by a series of Schr\"{o}dinger equations:
\begin{equation}
i\hbar \frac{\partial\psi_i^{probe}(x_{probe})}{\partial t} =  (-\frac{\hbar^2}{2 m_{probe}} + V(x_{probe} - x_{test}))\psi_i^{probe} \label{eq:branches1}
\end{equation}
and 
\begin{equation}
i\hbar \frac{\partial\psi_i^{test}(x_{test})}{\partial t} =  (-\frac{\hbar^2}{2 m_{test}} + V(x_{test} - x_{probe}))\psi_i^{test},
\label{eq:branches2}
\end{equation}
where $V(x)$ is a linear potential.
Before interaction, there is just one probe wave function corresponding to all different segments (equation \ref{eq:train}), but at arbitrary time $t$ generally speaking all $\psi_i^{probe}$ are different. 
After some time interval $t_D$, different $\psi_{tes}$ corresponding to equations \ref{eq:branches1} and \ref{eq:branches2} become almost orthogonal to each other. We can estimate this time interval using equations \ref{eq:Ehrenfest1}-\ref{eq:Ehrenfest4} for the averages
\begin{equation}
<x^{probe}_i(0)> \to <x^{probe}_i(t)> = <x^{probe}_i(0)> + \frac{F_i}{m_p} \frac{t^2}{2},
\end{equation}
where $F$ is the classical force acting between the particles. We define the decoherence time $t_{D}$ such that after $t_{D}$, $\psi_i^{probe}$ become almost orthogonal to each other. From equation \ref{eq:powerpotential}, this means that the distance between two consecutive averages will equal the width of the probe functions. This condition reads:
\begin{equation}
<x^{probe}_{i}(t)> - <x^{probe}_{i-1}(t)> = \sigma^{probe},
\end{equation}
For a force:
\begin{equation}
F=\pm\frac{q^2}{4 \pi \epsilon_0 r^2}, \label{eq:CoulombForce},
\end{equation}
we then get for the decoherence time interval:
\begin{equation}
t_{D} = \sqrt{\frac{2 \sigma_p m^{probe}}{F_i - F_{i-1}}} \label{eq:dec}
\end{equation}
Previously, we introduced parameter $\omega$ representing the degree of linearity of the potential. The width of a segment is given by:
\begin{eqnarray}
\sigma^{test}_{i-1} = \omega r_{i-1}\\
\sigma^{test}_{i-1} = r_i-r_{i-1}
\end{eqnarray}
Hence:
\begin{equation}
r_i - r_{i-1} = \omega r_{i-1}
\end{equation}
The final expression for the decoherence time interval is:
\begin{equation}
t_{D} = \sqrt{\frac{8 \pi \epsilon_0 \sigma_p m^{probe} (r_i^2 - r_{i-1}^2)}{q^2}} = \sqrt{\frac{8 \pi \epsilon_0 \sigma_p m^{probe} r_i^2 (2\omega + \omega^2)} {q^2}}\label{eq:decCoulomb}
\end{equation}

For an electron hitting a proton separated by a distance of 1 \text{\AA} (an order for the size of an atom) assuming the width of the 1 \text{\AA} for the atom, and taking $\omega$ as 0.01, we get $t_D \sim 10^{-14}s$. For slow electrons, for which time to pass an atom is longer than the $t_D$, the atom effectively measures the coordinates of the electron. However, for fast electrons, system doesn't have time to decohere.

In reality, our testing and probe systems are coupled to the thermal bath which makes the task of calculating decoherence time harder. The probe in this case will correspond to, say, a solid state consisting of multiple atoms. Interaction with each segment will shift $N$-atoms wave function. Since there are so many degrees of freedom, the orthogonality of $\psi_i^{probe}$ will be achieved much faster (see equation \ref{eq:scalarProduct}). It is beyond the scope of this paper to calculate decorehence times in this case. It was done say in \cite{bib:Zurek}. 
\section{Evolution of classical systems}
\label{sec:ev_classical}
Now, as we know that composite systems, that are not entangled with anything, behave as elementary objects in potential, we can consider their evolution. Suppose, we measure a coordinate of such a system. How does this system evolve in time? For:
\begin{equation}
\psi(x) \sim \exp(-\frac{(x-<x>)^2}{2 \sigma_x^2}) \label{eq:FT}
\end{equation}
in the momentum space, the corresponding wave function is:
\begin{equation}
\phi(p) \sim \exp(-\frac{(p-<p>)^2 \sigma_x^2}{2 \hbar^2}) \exp(-\frac{<x>^2}{2 \sigma_x^2})\label{eq:classicalevolution}
\end{equation}
Solving Schr\"{o}dinger equation, it is easy to show that the the width of the wave function evolves as:
\begin{equation}
\frac{\sigma^2_x(t)}{\sigma_0^2} = \sqrt{1+\left(\frac{\hbar t}{2 m \sigma_0^2} \right)^2} \label{eq:widthClassical}
\end{equation}
The characteristic time interval for the wave function to double its size therefore is:
\begin{equation}
t \sim \frac{m \sigma^2_0}{\hbar}
\end{equation}
This time is proportional to the mass of the system, and for the mass of 1 $kg$, and $\sigma$ of 1 \AA, it equals $10^{14} s$. During this time interval, the system will be measured again many times. Since the width of the wave packets almost doesn't change between the consequent measurements, we can assume that the macro-object follows classical trajectories. For an electron however, this time interval is $10^{-15}s$, therefore its evolution cannot be described using classical trajectories alone. 

For a hydrogen atom, the average time between interactions for $T \sim 100K$ $\sim10^{-12}s$ roughly equals the characteristic time to double its size. However, for smaller temperatures, the time between interactions is much larger leading to larger width in equation \ref{eq:widthClassical} and emergence of collective phenomena.

\section{Apparatus of an intermediate size}
What happens if we start scaling down our apparatus? The number of degrees of freedom $N$ also decreases, and at some point equation \ref{eq:orthogonality} stops being true. In this case, the branches in equation \ref{eq:systemAfterMeas} will become correlated.

\begin{figure}
\centering
\includegraphics[scale=0.5]{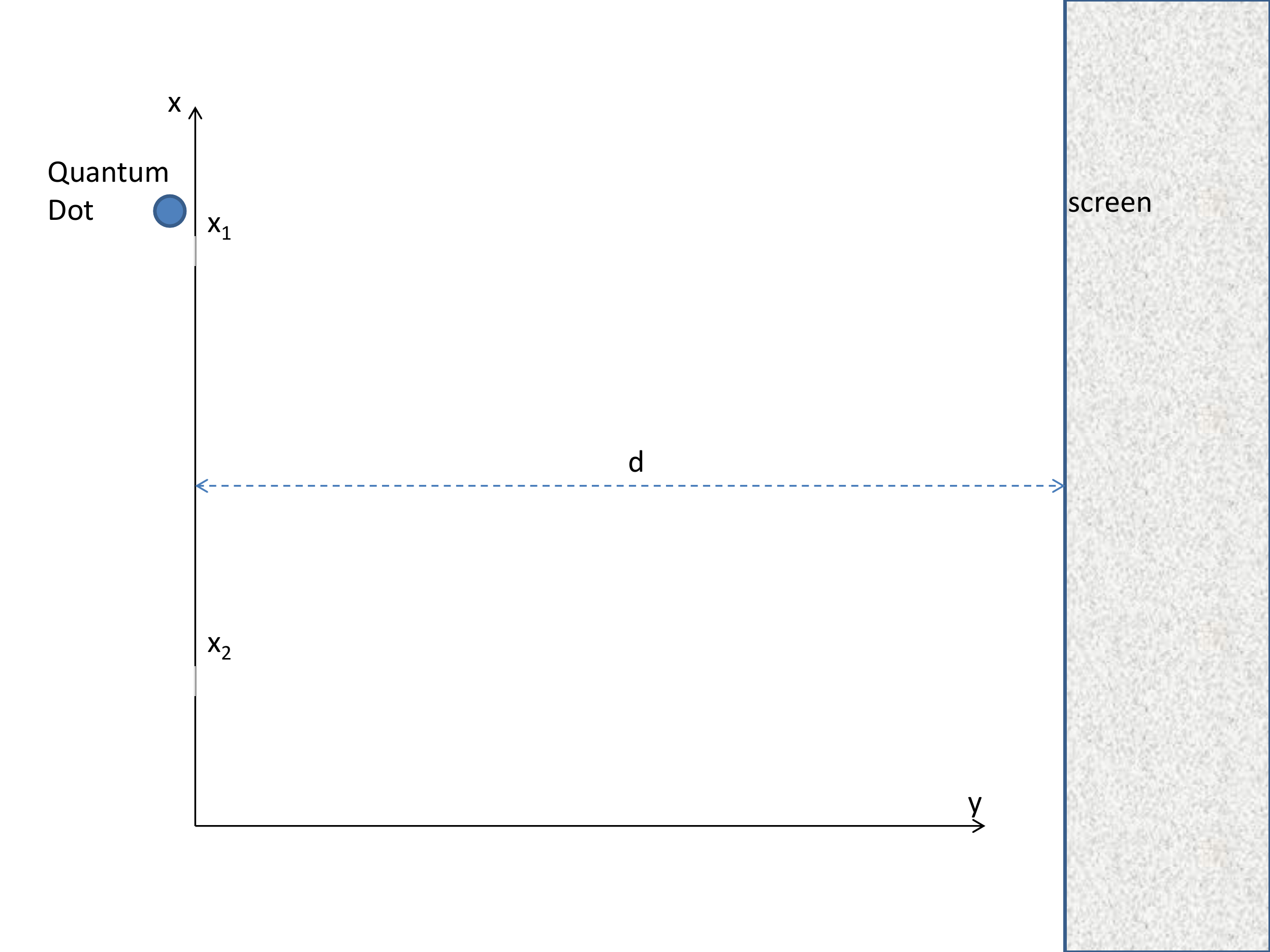}
\caption{Experimental setup for the double slit experiment with a quantum dot near one of the slits.}
\label{fig:DoubleSlit}
\end{figure}

Let's consider a double-slit experiment with a quantum dot near one of the slits (Fig. \ref{fig:DoubleSlit}). 
When the interaction of the electron with a quantum dot is negligibly weak, the electrons create a diffraction pattern on the screen. When the quantum dot is large, two orthogonal branches are generated, and there is no interference on the screen, the pattern resembles one produced by classical objects. For the quantum dot of an intermediate size, there is some non-zero correlation between the two states of the quantum dot. The strength of the diffraction term will depend on this correlation, and be maximum when there is no quantum dot. The state vector of the system electron-Quantum Dot located near the splits is given by:
\begin{equation}
|e, QD>_{splits} = \frac{1}{\sqrt{2}}(\ket{e_1}\ket{QD_1} + \ket{e_2}\ket{QD_2}), 
\end{equation}
where $QD$ is the quantum dot, and indices $1$ and $2$ correspond to the branches, so that $\ket{e_{1/2}}$ are the state of electron flying through the first or the second slit, and $\ket{QD_{1/2}}$ are corresponding states of the Quantum Dot. Also $\bra{e_1}\ket{x} = \frac{1}{\sqrt{2\pi}\sigma_1}\exp(-\frac{(x-x_1)^2}{2\sigma_1^2})$ and $\bra{e_2}\ket{x} = \frac{1}{\sqrt{2\pi}\sigma_2}\exp(-\frac{(x-x_2)^2}{2\sigma_2^2})$, where $x$ is the axis along the plane with two slits. Using equation \ref{eq:FT}, $\sigma_1 = \sigma_2 = \sigma_0$, introducing $k_x = \frac{p_x}{\hbar}$; $k_x^0 = <k_x>; \sigma_k = \frac{1}{\sigma_0}$, and Fourier transforming $\bra{e_1}\ket{x}, \bra{e_2}\ket{x}$, we obtain:
\begin{equation}
\psi(k_x, x_1^0) = \exp(-\frac{(k_x-k_x^0)^2}{2 \sigma_k^2})\exp(-\frac{(x_1^0)^2}{2 \sigma_x^2}),
\end{equation}
\begin{equation}
\psi(k_x, x_2^0) = \exp(-\frac{(k_x-k_x^0)^2}{2 \sigma_k^2})\exp(-\frac{(x_2^0)^2}{2 \sigma_x^2}),
\end{equation}
where $x_1^0$ and $x_2^0$ are the coordinates of the two slits.

As we have already found the operator of evolution, we use Schr\"{o}dinger equation with no potential to get the exact form of the wave functions hitting the screen. When traveling to the screen, the wave functions widen and get multiplied by a phase. The $x$ component of density of an electron when it passes the slits is given by: 
\begin{multline}
|\bra{e, QD}\ket{x}|^2_{screen} = \frac{1}{4\pi|\Delta|}|(\exp(-\frac{(x-x_1^0)^2}{2\Delta}) |QD_1> + \\ \exp(-\frac{(x-x_2^0)^2}{2\Delta})|QD_2>) |^2, 
\end{multline}
where $\Delta = \sigma_0^2$.
Solving Schr\"{o}dinger equation, we get at time $t$:
\begin{multline}
|\bra{e, QD}\ket{x}|^2_{screen} =
\frac{1}{4\pi|\Delta(t)|} |(\exp(-\frac{(x-x_1^0)^2}{2\Delta(t)})|QD_1> + \\ \exp(-\frac{(x-x_2^0)^2}{2\Delta(t)})|QD_2>) |^2, \label{eq:eltd}
\end{multline}
where $\Delta(t) = \sigma_0^2 + i\frac{\hbar}{2m} t$.
For an electron, as we discussed previously, the time it takes to double its width is $10^{-15} s$. Let us assume that the distance between the screen and the slits plane is $~1 m$. Even if the speed of the electron was the speed of light, the electron would have traveled only for $10^{-8} s \gg \frac{\sigma_0^2 m}{\hbar} \sim 10^{-15} s$, therefore we can safely assume that $\frac{\hbar}{2m} t \gg \sigma_0^2$ and $\Delta(t) \sim  i\frac{\hbar}{2m} t$.
After some simplifications, we get:
\begin{equation}
|<e, QD|x>|^2_{screen} = 2 + \\
\frac{1}{4\pi|\Delta(t)|} (\exp(-\frac{2 i m (x_1^0 - x_2^0)}{\hbar t } x)\bra{QD_1}\ket{QD_2} + H.C.)
\end{equation}
Defining $d$ as the distance between the slits and the screen, $k_y^0 = \frac{2 m d} {\hbar t}$ and choosing the origin of $x$ coordinates in the middle between the slits so that $x_1^0 = -x_2^0 = x^0$, we finally obtain for intensities since they are proportional to probabilities:
\begin{equation}
I(x) = 2 I_{1,2}(x) + I_{1,2} (\exp(ik_y^0(x_2^0-x_1^0)\frac{x}{d}) <QD_1|QD_2> + H.C.),
\end{equation}
where $I_{1,2} = I_1 = I_2$ is an intensity of a spot on the screen left by an electron if it passes through only one of the slits. When $\bra{QD_1}\ket{QD_2} \sim 0$, intensities add up, and when $|<QD_1|QD_2>| \sim 1$, the diffraction pattern is formed:
\begin{equation}
I(x) = 2 I_{1,2}(x) + 2 I_{1,2} \cos(k_y^0(x^0_2-x^0_1)\frac{x}{d}),
\end{equation}

Generally speaking, the apparatuses can be loosely split into four categories:
\begin{enumerate}
\item Macroscopic apparatuses. They branch the state vectors of the composite system.
\item Intermediate size apparatuses. The branches they created are slightly correlated and the evolution of the system can be traced using Schr\"{o}dinger equation.
\item Observers. Observers are apparatuses that try to observe the system they are entangled with. Since there is no way to disentangle the state vector of observer and the system, we are forced to conclude that each branch has its own observer and effectively collapses the state vector of the composite system. Any observer is a macroscopic apparatus, so an observation always effectively collapses the wave function. 
\item Observers of an intermediate size. The possibility to see a superposition would be possible only for an intermediate size observer who can analyze what he sees. All our senses are macroscopic so at present, such observers don't exist.
\end{enumerate}

\section{Conclusion}
We tried to build QM starting from CM and generalizing it until the final model is consistent. We managed to justify the following axioms and naturally arrive at the following important notions.
\begin{enumerate}
\item Complexity of the wave function.
\item Various representations and state vectors.
\item Special place of the momentum operator in QM.
\item Space spanned by non-interacting, interacting particles and weekly interacting particles.
\item Operators corresponding to measurement apparatuses.
\item Collapse and Quantum postulate.
\item Schr\"{o}dinger equation.
\item Possibility to describe composite objects with state vectors.
\item Narrowness and stability of wave function for classical objects.
\end{enumerate}
The rest of the results in QM are derived from these basic set of principles in the textbooks.

\section*{Acknowledgement}
I want to express my appreciation to Justin Gagnon, PhD and Vadim Asnin, PhD for comments that greatly improved the manuscript.

\end{document}